\documentclass[useAMS,usenatbib]{mn2e}
\usepackage{graphicx}
\usepackage{amsmath}

\begin{document}

\title{Simulations of bent-double radio sources in galaxy groups}

\author[B. J. Morsony et al.]{Brian J. Morsony,$^{1,2}$\thanks{Email:morsony@astro.wisc.edu} Jacob J. Miller,$^{1,3}$ Sebastian Heinz,$^1$ 
Emily Freeland,$^{4,5}$ \and Eric Wilcots,$^{1}$ 
Marcus Br\"uggen,$^{6,7}$ and Mateusz Ruszkowski$^{8,9}$\\
$^1${Department of Astronomy, University of
Wisconsin-Madison, 2535 Sterling Hall, 475 N. Charter Street, Madison WI
53706-1582, USA}\\
$^2${NSF Astronomy and Astrophysics Postdoctoral Fellow}\\
$^3${APS Department, University of Colorado, Boulder, UCB 391, Boulder, CO 80309, USA}\\
$^4${Dept. of Physics and Astronomy, Texas A\&M University, College Station, TX, 77843-4242}\\
$^5${Department of Astronomy \& Oskar Klein Centre, AlbaNova, Stockholm University, SE-106 91 Stockholm, Sweden}\\
$^6${Jacobs University Bremen, Campus Ring 1, 28759 Bremen, Germany}\\
$^7${Hamburger Sternwarte, University of Hamburg, Gojenbergsweg 112, 21029 Hamburg, Germany}\\
$^8${Department of Astronomy, The University of Michigan, 500 Church Street, Ann Arbor, MI 48109, USA}\\
$^9${The Michigan Center for Theoretical Physics, 3444 Randall Lab, 450 Church St, Ann Arbor, MI 48109, USA}}

\maketitle

\begin{abstract}

Bent-double radio sources have been used as a probe to measure the density of intergalactic gas in galaxy groups.
We carry out a series of high-resolution, 3-dimensional simulations of AGN jets moving through an external medium with a constant density in order to develop a general formula for the radius of curvature of the jets, and to determine how accurately the density of the intra-group medium (IGM) can be measured.  
Our simulations produce curved jets ending in bright radio lobes with an extended trail of low surface brightness radio emission.  The radius of curvature of the jets varies with time by only about $25\%$.
The radio trail seen in our simulations is typically not detected in known sources, but may be detectable in lower resolution radio observations.  The length of this tail can be used to determine the age of the AGN.
We also use our simulation data to derive a formula for the kinetic luminosity of observed jets in terms of the radius of curvature and jet pressure.
In characterizing how well observations can measure the IGM density, we find that the limited resolution of typical radio observations leads to a systematic under-estimate of the IGM density of about $50\%$.  The unknown angles between the observer and the direction of jet propagation and direction of AGN motion through the IGM leads to an uncertainty of about $\pm50\%$ in estimates of the IGM density.
Previous conclusions drawn using these sources, indicating that galaxy groups contain significant reservoirs of baryons
in their IGM, are still valid when considering this level of uncertainty.
In addition, we model the X-ray emission expected from bent-double radio sources.
We find that known sources in reasonably dense environments should be detectable in $\sim 100$~ks {\it Chandra} observations.  
X-ray observations of these sources would place constraints on the IGM density and AGN velocity that are complementary to radio observations.

\end{abstract}

\section{Introduction}

%
%

Observations of the local universe support the conclusions drawn by the hierarchical theory of structure formation that the majority of galaxies reside in groups, dynamically-bound systems which span a wide range of properties \citep{tully87}.  The intra-group medium (IGM) contained within these groups likely contains a significant fraction of the baryonic content of the universe \citep{Fuk98}.
Observations measuring the baryon content of the local universe account for approximately a third of the baryon density observed at high redshift ($z = 2-4$) in the form of stars, cold gas, and hot X-ray-emitting gas \citep{Fuk04,Stocke04}.
Numerical simulations predict a significant fraction of the ``missing'' baryons, around $40\%$ of the total baryonic content of the local universe, may be contained in the warm-hot intergalactic medium (WHIM) with a temperature range of $T \sim 10^5-10^7$~K \citep{Cen06,Dave01}.

The primary observational tool for studying the WHIM is currently absorption-line spectroscopy of low-redshift quasars \citep{Nara10}.  These observations are limited to groups falling along the line of sight of a sufficiently bright quasar, and density measurements depend on estimates of the systems' spatial extent, metallicity and ionizing fraction, yielding total IGM density measurements on the order $n \sim 10^{-4}-10^{-5}$~cm$^{-3}$ \citep{Pisano04}.
X-ray measurements are inherently limited to higher temperature groups, generally containing at least one early-type galaxy.  A dynamical mass can be estimated by making an assumption about the geometrical distribution of X-ray emitting gas in the group.  However, measurements of X-ray surface brightness drop below measurable levels well within the virial radius, requiring additional assumptions in order to extrapolate the total mass estimate of the group \citep{Mulchaey00}.

Observations of bent-double radio sources in galaxy groups serve as a valuable method of measuring IGM densities, and recently this method has been used to find total IGM densities in groups of $n \sim 2\times10^{-4} - 3\times10^{-3}$~cm$^{-3}$ \citep{Free08,Free10,Free11},
higher than the  density found from absorption-line measurements.
These measurements rely on a set of assumed physical parameters including viewing angle, AGN proper motion, and kinematic luminosity.  These assumptions are largely independent of those required for UV and X-ray density estimates, making this analysis complementary to existing methods.  Bent-double radio sources also probe the density of the entire IGM, rather than just one temperature phase.

Ram pressure resulting from the proper motion of these double-lobed radio sources through the IGM sweeps back the bipolar jets, producing the distinctive bent-double radio tails noted in many radio observations \citep{Miley72}.
Prior to \citet{Burns87}, it was believed that the conditions on ambient IGM density and galaxy velocities necessary to produce these bent-double radio tails could be found only in large, rich clusters of galaxies.  However, surveys have found a significant number of bent-double sources in lower mass galaxy groups \citep{Venk94,Doe95,Blanton01}.
\citet{Ekers78} was among the first to identify bent-double radio sources as a possible density probe of the IGM, observing IGM densities in the range $n \sim 3 \times 10^{-4}-6 \times 10^{-3}$~cm$^{-3}$ in the galaxy groups NGC~6109 and NGC~6137.  These early estimates have been corroborated by more recent observations of head-tail sources in galaxy groups and poor clusters \citep{Free08,Free11}.  If these numbers are representative of all galaxy groups, a significant fraction of the local universe's baryon content would reside within the IGM.

Analytic modeling of bent-double radio sources has found that the radius of curvature at the most bent part of the jet can be described by an equation of the form 

\begin{eqnarray}
\frac{R}{h} = \frac{P_{\rm jet}}{P_{\rm ram}}
\label{eqn:r_over_h}
\end{eqnarray}

\noindent which balances internal and external pressure gradients as a ratio of the radius of curvature $R$ and the scale height $h$ across which the pressure difference acts \citep{Begel79,Jones79,Burns80,Odea85}.  In \citet{Begel79}, $h$ is taken to be the diameter of the jet, whereas in Jones \& Owen, $h$ is the scale height of the ISM within the host galaxy.  We use the diameter of the jet for $h$.  We use this relationship as an analytic model to compare to our simulation results.

%
%

We carry out a series of numerical simulations of bent-double radio sources to determine values for $R$ and $h$ in terms of initial model parameters.  In \S 2, we describe the numerical methods and initial conditions used in our simulations.  In \S 3, we describe the results of our simulations and our methods for measuring $R$ and $h$.  In \S 4, we use our simulations to quantify the errors on density estimates from observations of bent-double radio sources.  We also develop a formula to estimate the kinetic luminosity of observed jets, and we make predictions for the X-ray detectability of radio sources.  In \S 5, we summarize our results.



\section{Technical description}

\subsection{Code description}

Simulations are carried out using the FLASH 2.4 hydrodynamics code \citep{Fryxell00}, which is a modular, block-structure adaptive mesh code. It solves the Riemann problem on a three-dimensional Cartesian grid using the piecewise-parabolic method. 
%
%
Gas is modeled as having a uniform adiabatic index of $\gamma = 5/3$.
%

\subsection{The Jet Nozzle}

In order to simulate the injection of collimated, supersonic jets into the grid, we employ a numerical ``nozzle'', as first developed and described in \citet{Heinz06}: an internal inflow boundary of cylindrical shape placed at the location of the AGN, injecting fluid with a prescribed energy, mass, and momentum flux to match the parameters we choose for the jet.
For reasons of numerical stability, we impose a slow lateral outflow with low mass flux in order to avoid complete evacuation of zones immediately adjacent to the nozzle due to the large velocity divergence at the nozzle. The injection of energy and mass due to this correction is negligible.

We model unresolved dynamical instabilities near the base of the jet by imposing a random-walk jitter on jet axis confined to a $5\degr$ half-opening angle. This is necessary to model the `dentist's drill' effect of \citet{Scheuer82}.
%
The time for the jet to change direction is slow compared to time for jet material to reach the end of the jet, so the jitter does not significantly change the bending of the jet, aside from symmetry breaking.  
It does, however, change the shape of the initial cocoon created around the jet (see section~\ref{sect:results}) by spreading energy over a wider average angle.

We chose to inject the jet at an internal Mach number of 10 in most simulations. 
%
For computational feasibility, we chose a jet velocity of $v_{\rm jet} = 3 \times 10^9$~cm/s for most simulations. 
%
For a typical case (i.e. model .25E), this results in a jet density of $n_{\rm jet} = 1.82\times10^{-5}$~cm$^{-3}$, pressure of $p_{\rm jet} = 1.64\times10^{-12}$~erg$/$cm$^{3}$ and temperature of $T_{\rm jet} = 6.53\times10^8$~K.
The jet is turned on initially and continues to inject material for the entire length of the simulation.

\subsection{Initial Conditions}

Our setup is similar to that used for X-ray binary jets in \citet{Yoon11}.  We place the jet nozzle in a moving medium inside a simulation box large enough that the boundaries never affect the simulation ($2.8$~Mpc on a side).  
We keep the location of the AGN fixed in space, letting the IGM stream by at velocity $-v_{\rm gal}$ perpendicular to the jet axis.  We can vary the velocity and density of the IGM, as well as the luminosity, velocity and internal Mach number of the jet.  
%
In all cases the IGM pressure is set to $2.76 \times 10^{-13}$~erg/cm$^3$, giving a typical IGM sound speed of $166$~km/s and temperature of $2\times10^6$~K.  
A list of parameters for all simulations is given in table~\ref{table:model_data}.

The simulations were carried out on a staggered mesh grid as described in \citet{Yoon11} in order to capture the large dynamic range required, and ensure that the nozzle diameter is resolved by 12 grid cells. For our normal scaling, the nozzle diameter is 2~kpc, with a maximum resolution for the standard model of about 0.175~kpc near the jet nozzle.  
The nozzle diameter for each simulation, d$_{\rm j}$, is listed in table~\ref{table:model_data}.  
In all cases, the maximum resolution is set such that the nozzle is resolved by 12 grid cells.
Short duration test simulations at higher resolution were carried out and produced similar results.

One simulations was for a conical jet with an initial opening angle, rather than a purely parallel inflow.  This model, .015E\_theta20, has identical parameters to model .015E, except that the nozzle is 4 times smaller (and the maximum resolution 4 times greater) and the inflowing material is spread over a $20\degr$ half-opening angle.  This model was run as a direct comparison to model .015E, based on the analytic model in section~\ref{sect:analytic}.  We use cylindrical (parallel) jet injection for our other models because this allows us to use a much lower resolution.




\begin{table*}
\centering
\begin{minipage}{140mm}
\caption{Model Data}
\begin{tabular}{@{}lllllllll@{}}
\hline
Model & $L_{\rm jet}$ & $n_{\rm IGM}$ & $v_{\rm gal}$ & $v_{\rm jet}/c$ & M & R & h & d$_{\rm j}^a$ \\
 ~ & $(10^{44}$~erg/s) & (cm$^{-3}$) & (km/s) & ~ & ~ & (kpc) & (kpc) & (kpc) \\
\hline

.015E\_theta20 & $0.015625$ & $1\times10^{-3}$ & $1000$ & $0.1$ & $10$ & $8.6\pm0.8$ & $0.44\pm.02$ & 0.125$^b$ \\
.015E &        $0.015625$ & $1\times10^{-3}$ & $1000$ & $0.1$ & $10$ & $9.4\pm1.4$ & $0.49\pm.01$ & 0.5 \\
.062E &        $0.0625$ & $1\times10^{-3}$ & $1000$ & $0.1$ & $10$ & $18.4\pm5.2$ & $0.91\pm.03$ & 1.0 \\
.25E &         $0.25$ & $1\times10^{-3}$ & $1000$ & $0.1$ & $10$ & $35.5\pm7.9$ & $2.06\pm.07$ & 2.0 \\
1E &           $1.0$ & $1\times10^{-3}$ & $1000$ & $0.1$ & $10$ & $76.4\pm15.3$ & $3.87\pm.14$ & 4.0 \\
.062E\_.5vel &  $0.0625$ & $1\times10^{-3}$ & $500$ & $0.1$ & $10$ & $35.4\pm6.9$ & $2.05\pm.06$ & 2.0 \\
.015E\_.25vel & $0.015625$ & $1\times10^{-3}$ & $250$ & $0.1$ & $10$ & $38.9\pm10.8$ & $1.94\pm.07$ & 2.0 \\
1E\_4n &        $1.0$ & $4\times10^{-3}$ & $1000$ & $0.1$ & $10$ & $35.5\pm12.2$ & $2.09\pm.10$ & 2.0 \\
.062E\_.25jvel &   $0.0625$ & $1\times10^{-3}$ & $1000$ & $0.025$ & $10$ & $30.4\pm3.3$ & $2.41\pm.14$ & 2.0 \\
.25E\_4M &      $0.25$ & $1\times10^{-3}$ & $1000$ & $0.1$ & $40$ & $40.1\pm6.0$ & $1.89\pm.05$ & 2.0 \\

\hline
\label{table:model_data}
\end{tabular}

$^a$ Nozzle diameter at injection

$^b$ This simulations has a half-opening angle of $20\degr$ at injection, rather than a parallel inflow
\end{minipage}
\end{table*}


%


\section{Simulation Results }
\label{sect:results}


When the jet initially turns on, it drives a shock into the IGM, creating a rapidly expanding cocoon.  
The expansion velocity of the cocoon is initially faster than $v_{\rm gal}$, so the AGN remains inside the cocoon and the jet stays fairly straight.
As the cocoon expands it decelerates, but $v_{\rm gal}$ remains constant.
Eventually, the AGN moves outside the initial cocoon and a bow shock is created around the AGN jets.  
The pressure gradient from this shock bends the jets backwards, creating the characteristic curved structure.
At some point along the jet, the flow becomes unstable and breaks up, creating bright radio lobes at the ends of the jets.
The swept-back jet material creates a long tail of diffuse jet material reaching back to the cocoon created around the initial position of the AGN.

We create synthetic radio images by assuming that the jet material, followed with a tracer fluid, consists of a hot plasma of relativistic particles with magnetic fields in equipartition with thermal pressure.  
Energy lose from synchrotron cooling is not included.

\begin{figure*}
\includegraphics[scale=.8]{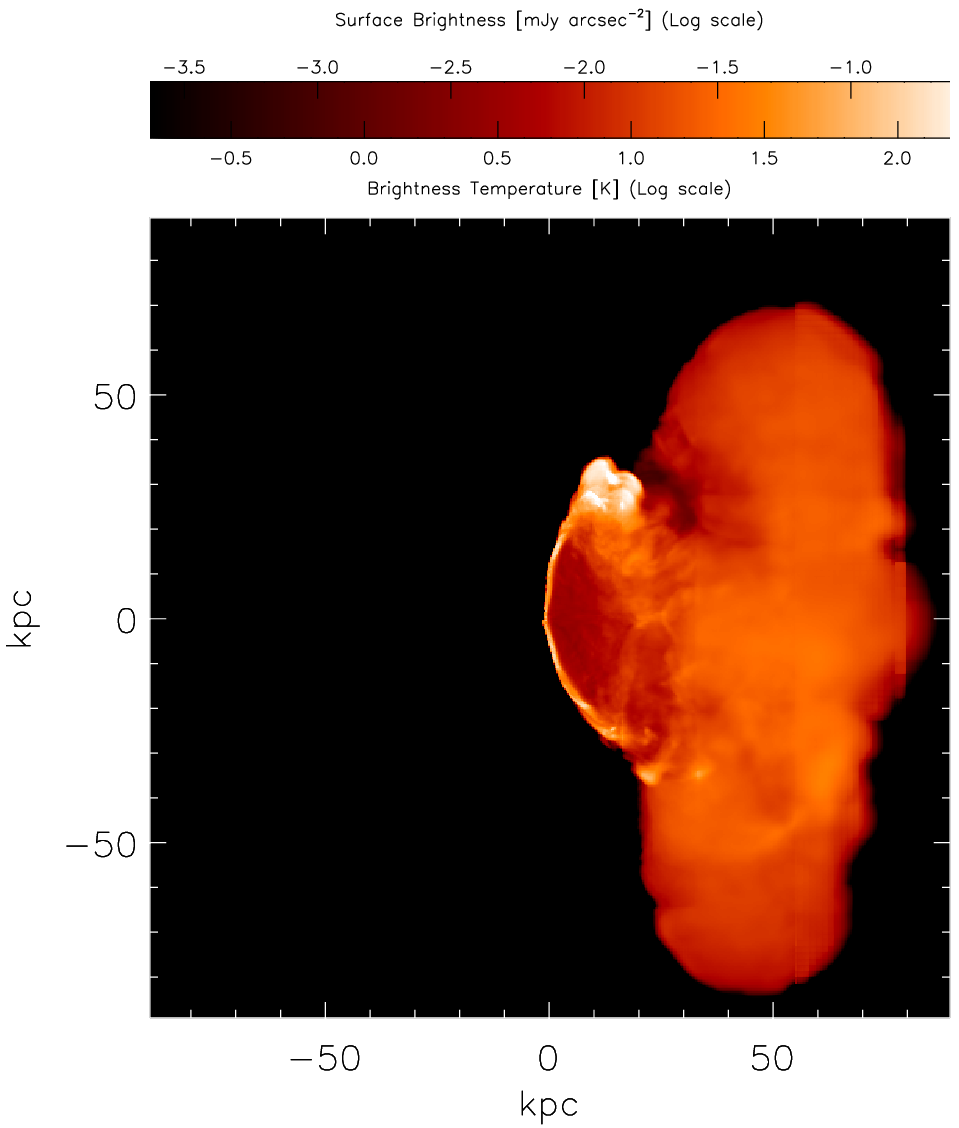}
\includegraphics[scale=.8]{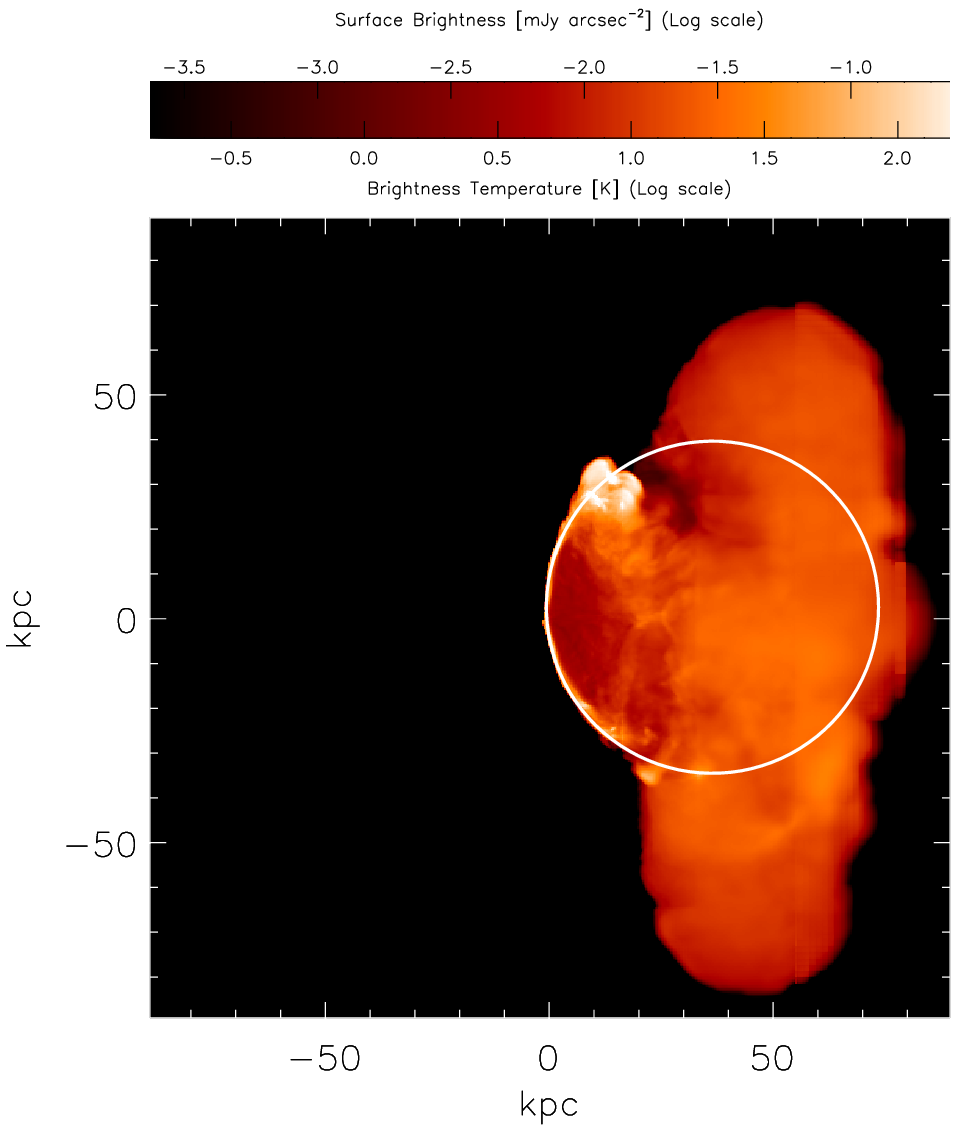}
\caption{
Left: Simulated radio emission from model .25E at 50 Myr (log scale).  
Color bar shows intensity in mJy/arcsec$^2$ and brightness temperature at 1440 MHz.  
The image is $170$~kpc on a side.  
In addition to the two bright radio jets, there is diffuse synchrotron emission filling the cocoon created by the jets.  The surface brightness of this emission is $\approx10-100$ times fainter than the jets.  Right: Same as left panel, but overlaid with the best fit radius of curvature (white circle).  The radius is 37~kpc.
}
\label{fig:radio_example}
\end{figure*}

The left panel of Fig.~\ref{fig:radio_example} shows a synthetic radio image for simulation .25E at 50~Myr.  
In this image the AGN is moving relative to the IGM from right to left.  
The narrow curved jets and bright radio lobes can be clearly seen.  
The tail and cocoon produce diffuse radio emission with a surface brightness 1 to 2 orders of magnitude fainter than the jet and lobes.  This extended emission is typically not seen in bent-double radio sources in jets.
It is possible that this emission is resolved out in existing observations or that synchrotron cooling, which is not included in making our images, makes the emission at $1.4$~GHz too faint to observe.  Observations at lower resolution and/or lower frequencies may reveal the extended tail.  
It is possible that at least part of the tail is seen in source S7 in \citet{Free11}, which appears to have extremely thick jets relative to the radius of curvature.  However, this is not the only explanation for this source's appearance (see section~\ref{sect:viewing_angle}).

The length of the tail is $v_{\rm gal} \times t_{\rm AGN}$, where $t_{\rm AGN}$ is the amount of time the AGN has been active.
If there is an estimate for the AGN host galaxy velocity, finding the length of the tail would allow the amount of time the AGN has been active to be determined.
Note that in Fig.~\ref{fig:radio_example} the length of the tail (left to right) is less than the width of the tail (top to bottom), because the jets initially expand outwards faster than $v_{\rm gal}$.  
If the AGN remains active, the tail will continue to lengthen and eventually become longer than it is wide.
%


\subsection{Radius of Curvature \label{sect:measure_r}}

The right panel of Fig.~\ref{fig:radio_example} shows the same image with a circle overlaid with $R = 37$~kpc, the best fit radius of curvature for this image.  The circle traces both the upper and lower jet and passes through the bright radio lobes where each jet breaks up.  
Note that the upper lobe is significantly brighter than the lower lobe at this time.  The brightness of the lobes and the curvature of the two jets varies with time due to instabilities in the jet propagation and the small random changes in jet direction that we introduce.

The radius of curvature is determined by an automated fitting procedure.  We start with a synthetic radio image and consider only the region $28$~kpc above and below of the AGN, the approximate extent of the jets.  Because the upper and lower jets and lobes can differ significantly in brightness, we find the maximum brightness of any point in each half of the image and exclude points with less than $10\%$ of this brightness.  For each slice along the jet, we then determine the horizontal location of the jet by taking an intensity weighted average of the radio emission in that slice.  We then fit a circle to the jet locations weighted by the total radio intensity of the points in each slice.

\begin{figure*}
\includegraphics[scale=1.]{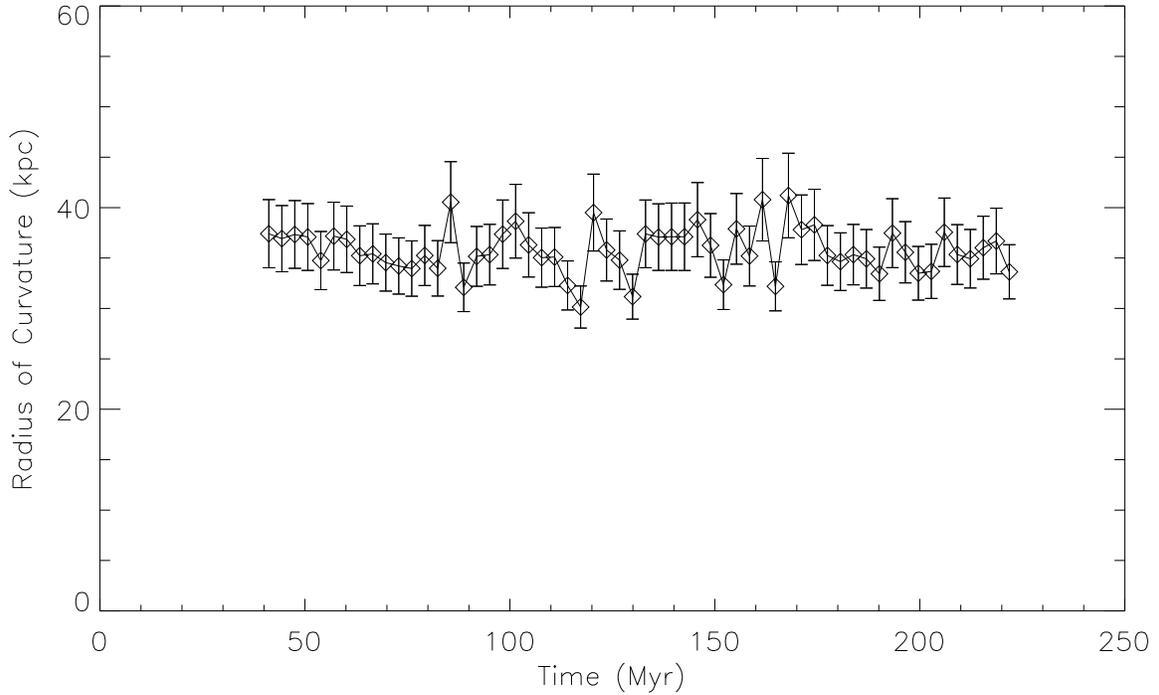}
\caption{
Best fit radius of curvature at different times for model .25E.  The radius varies about $15\%$ between measurements.  The overall mean radius is 35.5 kpc with an error of 7.9 kpc.
}
\label{fig:curve_fits_.25E}
\end{figure*}

To characterize how much fluctuations in the jet affect the radius of curvature, we use the above procedure to measure the radius of curvature at a series of times (after the jet curvature is well established) and compare the results.  
Fig.~\ref{fig:curve_fits_.25E} plots measured values of $R$ for simulation .25E from $40$~Myr to $220$~Myr.  The error bars represent the $1$-$\sigma$ error on the value of $R$ at each time, typically about $10\%$.  The values of $R$ are very consistent, with a scatter of about $25\%$.  The best-fit value of $R$ for simulation .25E is $35.5\pm7.9$~kpc, where the error takes into account both the error in individual fits and variations between different snapshots.  
The same procedure is applied to all of our simulations, with the results listed as $R$ in table~\ref{table:model_data}.  


\subsection{Jet Thickness}

A similar process is used to find the average jet thickness, listed as $h$ in table~\ref{table:model_data}.
The thickness typically has a very small variation, $\simeq 5\%$, and the accuracy is limited by the resolution of our simulations.
Because we are interested in the ``real'' value of the jet thickness, we determine it using the raw simulation data rather than the synthetic radio data.  We take a region of $\pm24$~kpc from the AGN and define points as being in the jet if at least $80\%$ of the material at that point was injected by the nozzle and it is moving with at least $80\%$ of the initial jet velocity.  
This excludes all material in the lobes and the cocoon around the jet.
We then determine the jet thickness perpendicular to both the jet direction and the motion of the AGN, henceforth the `z' direction, as this is the front along which the ram pressure acts.
For each slice through the jet we take the average thickness for points in the jet in the `z' direction.  
We then take the average of this thickness in all slices to get a value for $h$ at that time, with the standard deviation of the average being the uncertainty at that time.  
To get the values of $h$ in table~\ref{table:model_data}, we then find the best-fit value of the thickness for all times considered, and the error takes into account both the error at individual times and the variation with time.


The overall result is that the radius measured at a particular time is accurate to within an error of 25\% of the `true' value for that particular combination of parameters.  The thickness is more consistent, with typically only a $5\%$ variation.  The variation in $R$ sets a lower limit on the accuracy of the IGM density estimated from observations of bent-double radio sources.


\section{Discussion}

%
%

\subsection{Analytic Fit for Radius of Curvature \label{sect:analytic}}

For an AGN moving supersonically relative to the IGM, the jets will curve due to the ram pressure of the incoming IGM.  At the point of maximum curvature, the ratio of the radius of curvature to the thickness of the jet will equal the ratio of the ram pressure of the jet to the external ram pressure \citep{Begel79,Jones79,Burns80}, i.e.:

\begin{eqnarray}
\frac{R}{h} = \frac{P_{\rm jet}}{\rho_{\rm IGM} v_{\rm gal}^2}
\label{eqn:r_over_h}
\end{eqnarray}

\noindent The ram pressure of the jet will be

\begin{eqnarray}
P_{\rm jet} = \frac{L_{\rm jet}}{\frac{\pi}{4} h^2 v_{\rm jet}}
\label{eqn:P_jet}
\end{eqnarray}

%
%

Rearranging eqn.~\ref{eqn:r_over_h} and eqn.~\ref{eqn:P_jet} gives a formula for $R$ in terms of $h$: 

\begin{eqnarray}
R = \frac{L_{\rm jet}}{\frac{\pi}{4} h v_{\rm jet} \rho_{\rm IGM} v_{\rm gal}^2}
\label{eqn:r_with_h}
\end{eqnarray}

\noindent However, the thickness of the jet is not constant and will be determined by the external pressure.  
Initially the lateral expansion of the jet will be ballistic because the component of the ram pressure perpendicular to the jet will be higher than the external pressure. 
However, as the thickness increases the internal pressure will drop until it is the same order as the cocoon (i.e. external) pressure.  
At this point, a recollimation shock will be driven into the jet, setting the thickness and providing an internal pressure that is equal to that of the cocoon.  Therefore, the thickness will be set such that the external pressure balances the perpendicular ram pressure.
For a jet with an initial half-opening angle of $\theta$, the average ram pressure of the jet perpendicular to direction of motion will be 
$P_{\rm jet,\perp} \simeq \frac{1}{2} P_{\rm jet} \sin^2 \theta$.  
Balancing this against the external pressure gives

\begin{eqnarray}
\rho_{\rm IGM} v_{\rm gal}^2 = \frac{1}{2} P_{\rm jet} \sin^2 \theta = \frac{1}{2} \frac{L_{\rm jet}}{\frac{\pi}{4} h^2 v_{\rm jet}} \sin^2 \theta
\label{eqn:p_perp}
\end{eqnarray}

\noindent Solving for h, we find 

\begin{eqnarray}
h = \frac{\sin \theta}{\sqrt 2} \left( \frac{L_{\rm jet}}{\frac{\pi}{4} v_{\rm jet} \rho_{\rm IGM} v_{\rm gal}^2} \right)^{1/2}
\label{eqn:h}
\end{eqnarray}

\noindent Using this value for $h$ in eqn.~\ref{eqn:r_with_h} gives us

\begin{eqnarray}
R = \frac{2 h}{\sin^2 \theta} = \frac{\sqrt 2}{\sin \theta} \left( \frac{L_{\rm jet}}{\frac{\pi}{4} v_{\rm jet} \rho_{\rm IGM} v_{\rm gal}^2} \right)^{1/2}
\label{eqn:R}
\end{eqnarray}

%
%
%
%
%
%

%
%

For our simulations with a cylindrical jet, we fix the nozzle size to h set by 

\begin{eqnarray}
h &=& 4\textrm{~kpc} \times \left(\frac{L_{\rm 44}}{n_{\rm -3} v_{\rm gal,1000}^2 v_{\rm jet,0.1}}\right)^{1/2} 
\label{eqn:h_for_parallel}
\end{eqnarray}

\noindent where $L_{\rm 44} = L_{\rm jet}/10^{44}$~erg/s, $n_{\rm -3} = n_{\rm IGM}/10^{-3}$~cm$^{-3}$, $v_{\rm gal,1000} = v_{\rm gal}/1000$~km/s, and $v_{\rm jet,0.1} = v_{\rm jet}/(0.1 c)$.  
This is the equivalent width of a jet with an initial opening angle of $\theta = 20\degr$.  From eqn.~\ref{eqn:R}, this would predict a value of $h/R = 1/17$.

We run one model, .015E\_theta20, with a small nozzle size and a $20\degr$ initial opening angle.  The measured values of $R$ and $h$ in talbe~\ref{table:model_data} are $8.6\pm0.8$~kpc and $0.44\pm0.2$~kpc, respectively, very close to the prdicted values of $8.5$~kpc and $0.49$~kpc given by equations~\ref{eqn:R} and \ref{eqn:h}.  The results are also very similar to model .015E, which has the same setup but with a cylindrical jet input and a $0.5$~kpc nozzle size.

Using the measured values of $R$ and $h$ in table~\ref{table:model_data} for all of our parallel jet models, we find that the value of $h$ stays very close to the nozzle size and the ratio of $h/R$ is about $1/18$, close the our analytic estimate of $1/17$.

Using the measured values of $R$ in table~\ref{table:model_data}, we find that the value of $R$ in terms of our model parameters is

\begin{eqnarray}
R &=& 5\textrm{~} \left(\frac{L_{\rm jet}}{\rho v_{\rm gal}^2 v_{\rm jet}}\right)^{1/2} \textrm{~cm} \nonumber \\
&=& 72 \textrm{~kpc} \times \left(\frac{L_{\rm 44}}{n_{\rm -3} v_{\rm gal,1000}^2 v_{\rm jet,0.1}}\right)^{1/2} 
\label{eqn:radius_fit}
\end{eqnarray}

%

\begin{figure*}
\includegraphics[scale=1.]{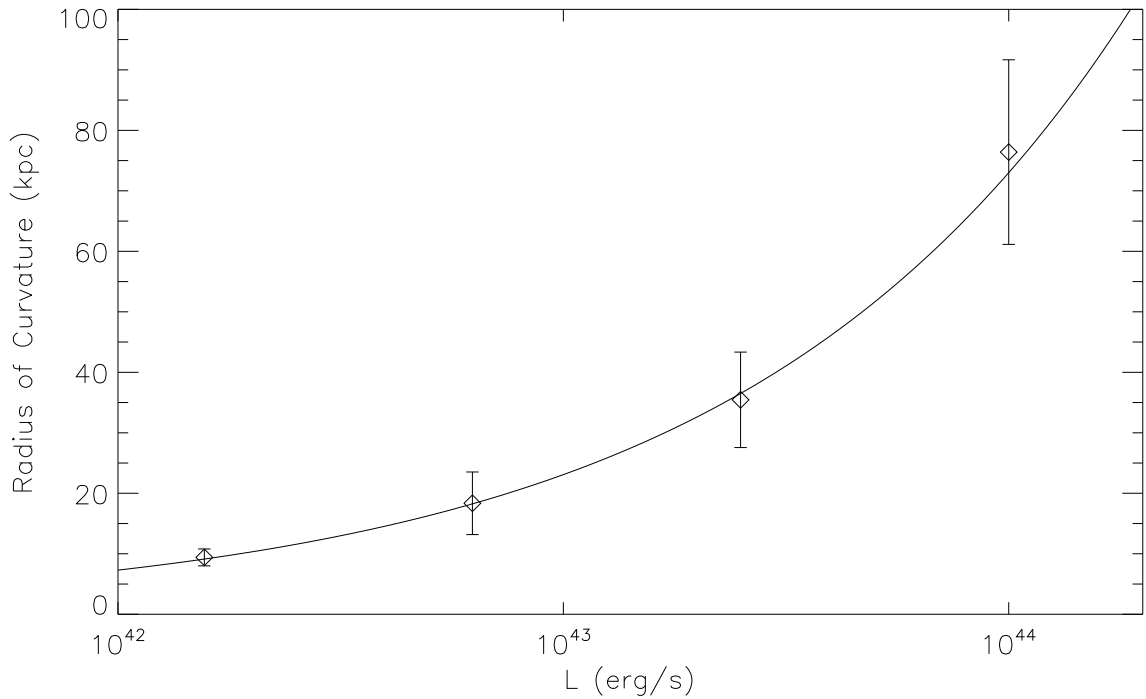}
\caption{
Radius of curvature vs. jet luminosity (diamonds) for models .015E, .062E, .25E and 1E.  These models differ only in jet luminosity, with all other parameters the same.  Error bars represent the overall error in measuring $R$.  Solid line is a plot of eqn.~\ref{eqn:radius_fit} for radius vs. luminosity.
}
\label{fig:curve_vs_energy}
\end{figure*}

Figure~\ref{fig:curve_vs_energy} plots the radius of curvature vs. jet luminosity for models .015E, .062E, .25E and 1E.  These models differ only in jet luminosity, with all other parameters the same.  Error bars represent the overall error in measuring $R$, as discussed in section~\ref{sect:measure_r}. The solid line is a plot of eqn.~\ref{eqn:radius_fit} for radius vs. luminosity and passes well within the error bars of all the four data points.

Equation~\ref{eqn:R} can also be used to find the kinetic luminosity of observed bent-double radio sources based on their measured radius of curvature and jet pressure.  
In \citet{Free11}, $P_{\rm jet}$ is assumed to be the minimum synchrotron pressure 

%
\begin{eqnarray}
P_{\rm jet} = P_{\rm min} = (2\pi)^{-3/7} \left( \frac{7}{12} \right) [c_{\rm 12} L_{\rm rad} (1+k) (\phi V)^{-1} ]^{4/7}
\label{eqn:pmin}
\end{eqnarray}

\noindent where $c_{\rm 12}$ is a constant that depends on the spectral index and frequency cutoffs \citep{Pacholczyk70}, $k$ is the ratio of relativistic proton to relativistic electron energy, $\phi$ is the volume filling factor, $V$ is the source volume, and $L_{\rm rad}$ is the radio luminosity of the jet at the point where the pressure is measured.
$P_{\rm jet}$ is measured in several slices along the jet, with the volume assumed to be proportional to the square of measured jet thickness, which is limited by the beam size of the observations.  Therefore, $P_{\rm min} \propto h^{-8/7}$ but is independent of the radius of curvature $R$ and the length of the jet.

%
%

Rearranging eqn.~\ref{eqn:R}, we find a jet luminosity of

\begin{eqnarray}
L_{\rm jet} &=& \frac{\pi}{8} R^2 \sin^2 \theta \rho_{\rm IGM} v_{\rm gal}^2 v_{\rm jet}  \nonumber \\
&=& 3.73 R_{\rm kpc}^2 \left( \frac{h}{R} \right) n_{\rm IGM,-3} v_{\rm gal,1000}^2 \beta_{\rm jet} \times 10^{42} \textrm{~erg/s}
\label{eqn:L_fit_dens}
\end{eqnarray}


\noindent where $R_{\rm kpc}$ is $R$ in kpc and $\beta_{\rm jet} = v_{\rm jet}/c$.  In terms of pressure $L_{\rm jet}$ is

\begin{eqnarray}
L_{\rm jet} &=& \frac{\pi}{4} R^2 \left( \frac{h}{R} \right)^2 P_{\rm min} v_{\rm jet}  \nonumber \\
&=& 2.24 h_{\rm kpc}^2 P_{\rm min,-11} \beta_{\rm jet} \times 10^{42} \textrm{~erg/s}
\label{eqn:L_fit_pres}
\end{eqnarray}


\noindent where $h_{\rm kpc}$ is $h$ in kpc. 

The jet kinetic luminosity calculated using this formula for the sources and measured values of $R$, $h$ and the synchrotron pressure $P_{\rm min}$ in \citet{Free11} are listed in table~\ref{table:sources_emily}.  
The values in table~\ref{table:sources_emily} are upper limits on the luminosity made with the assumptions that the observed value of $h$ is the true jet thickness and that the jet velocity is $v_{\rm jet} = c$.  
If the jet is narrower or slower, the luminosities will be smaller.
Note that the formula for $L_{\rm jet}$ is independent of the AGN velocity, so values can be found even for sources with unconstrained velocities.
The sources are all of order $10^{45}$~erg/s and the variation between sources is smaller than the variation in the total radio power at $1440$~MHz ($L_{\rm 1440}$).
$P_{\rm min}$ is proportional to $L_{\rm rad}^{4/7}$, which is the total radio emission at the point where $P_{\rm min}$ is measured.  $L_{\rm rad}$ therefore depends on the $1440$~MHz emission at that point and the model of the synchrotron spectrum used, but is not directly dependent on the total $1440$~MHz emission of the entire source.
$P_{\rm min}$ scales with jet thickness as $h^{-8/7}$, so $L_{\rm jet} \propto h^{6/7}$. 


Note that we assume the ram pressure of the jet, $P_{\rm jet}$, is accurately reflected by the minimum synchrotron pressure, $P_{\rm min}$, which is true only if the jet energy is dominated by relativistic electrons and magnetic fields.  
If the true jet ram pressure is higher, the estimates of both $n_{\rm IGM}$ and $L_{\rm jet}$ in table~\ref{table:sources_emily} will both be proportionally higher. 
Adding invisible components to the momentum flux of the jet only acts to increase the required IGM density.
On a similar note, in our simulations very little external material becomes entrained in the jets.
However, even if a significant amount of mass is entrained it will not change the momentum flux of the jet.
Therefore, the radius of curvature and inferred IGM density should not affected.
Also note that the equations in this section were derived assuming that the AGN host galaxy is moving supersonically relative to the IGM and that the jet velocity is supersonic relative to its internal sound speed.  These relations are not expected to hold if galaxies or jets are subsonic.


\begin{table*}
\centering
\begin{minipage}{180mm}
\caption{Jet Kinetic Luminosity}
\begin{tabular}{@{}llllllll@{}}
\hline
Source ID & $v_{\rm gal}^{a,b}$ & $P_{\rm min}^a$ & $n_{\rm IGM}^a$ & $h^a$ & $R_{\rm bend}^a$ & $L_{\rm 1440}^a$ & $L_{\rm jet}^c$ \\
 ~  & (km/s) & $10^{-11}$~erg~cm$^{-3}$ & (cm$^{-3}$) & (kpc) & (kpc) & (W Hz$^{-1}$) & (erg/s) \\
\hline

S1 & $430^{+170}_{\rm -35}$ & $0.9\pm0.2$ & $3\pm2 \times 10^{-3}$ & $23\pm1$ & $42\pm6$ & $1.6 \times 10^{25}$ & $1.1\pm0.2 \times 10^{45}$\\
S2 & $570\pm60^{d}$ & $0.6\pm0.2$ & $5\pm4 \times 10^{-4}$ & $30\pm4.5$ & $104\pm9$ & $1.88 \times 10^{25}$ & $1.2\pm0.5 \times 10^{45}$\\
S3 & $745^{+109}_{\rm -80}$ & $1.4\pm0.6$ & $2\pm1 \times 10^{-4}$ & $10\pm0.6$ & $141\pm19$ & $3 \times 10^{23}$ & $3.1\pm1.4 \times 10^{44}$\\
S4 & $950^{+210}_{\rm -140}$ & $1.7\pm0.3$ & $5\pm2 \times 10^{-4}$ & $22\pm0.7$ & $89\pm7$ & $9.5 \times 10^{24}$ & $1.5\pm0.3 \times 10^{45}$\\
S5 & Unconstrained & $0.4\pm0.1$ & $(70\pm21)/v^2$ & $38\pm3.8$ & $220\pm11$ & $6.4 \times 10^{24}$ & $1.3\pm0.4 \times 10^{45}$\\
S6 & Unconstrained & $0.6\pm0.1$ & $(156\pm48)/v^2$ & $19\pm3.1$ & $69\pm4$ & $9 \times 10^{23}$ & $4.8\pm1.4 \times 10^{44}$\\
S7 & $850^{+170}_{\rm -120}$ & $1.4\pm0.3$ & $2\pm1 \times 10^{-3}$ & $22\pm3.6$ & $18\pm4$ & $2 \times 10^{24}$ & $1.5\pm0.5 \times 10^{45}$\\

\hline
\label{table:sources_emily}
\end{tabular}

$^a$~from~\citet{Free11}.

$^b$~$v_{\rm gal}$ here is $\sqrt3$ times the group velocity dispersion, which is listed as $v_{\rm gal}$ in Table 1 of \citet{Free11}.

$^c$~Upper~limit,~assuming~$v_{\rm jet} = c$

$^d$~For S2, velocity is based on the difference in redshift between the source galaxy and the group, not the velocity dispersion.
\end{minipage}
\end{table*}


\subsection{Effects of Observational Resolution}

Although the jets are well resolved in our simulations, they are typically unresolved in radio observations.  From eqn.~\ref{eqn:r_over_h} and eqn.~\ref{eqn:pmin}, the density derived from observations will scale with jet thickness and radius of curvature as $n_{\rm IGM} \propto (h/R)^{-1/7} R^{-8/7}$.  The values of $h/R$ in our simulations (table~\ref{table:model_data}) range from about $1/13$ to $1/21$ with a typical value of about $1/18$.  
The observed ratio (table~\ref{table:sources_emily}) ranges from $h/R = 1/14$ for S3 to $h/R = 1/0.83$ for S7.  
Although the density scales weakly with jet thickness, if we assume a real value of $h/R = 1/17$, corresponding to an initial jet opening angle of $\theta = 20\degr$, the densities for observed sources would be between 3\% (S3) and 54\% (S7) higher.  
If a $\theta = 5\degr$ inital opening angle is assumed, the correction would be between 52\% (S3) and 128\% (S7)
For an under-resolved jet, the observed thickness will always be too high, and therefore the density derived will always be lower than the actual value.  
This is generally a fairly small correction, about 25\% for a typical source, but can be up to 50\% or more for sources with $h_{\rm obs} \sim R$.

\begin{figure*}
\includegraphics[scale=1.]{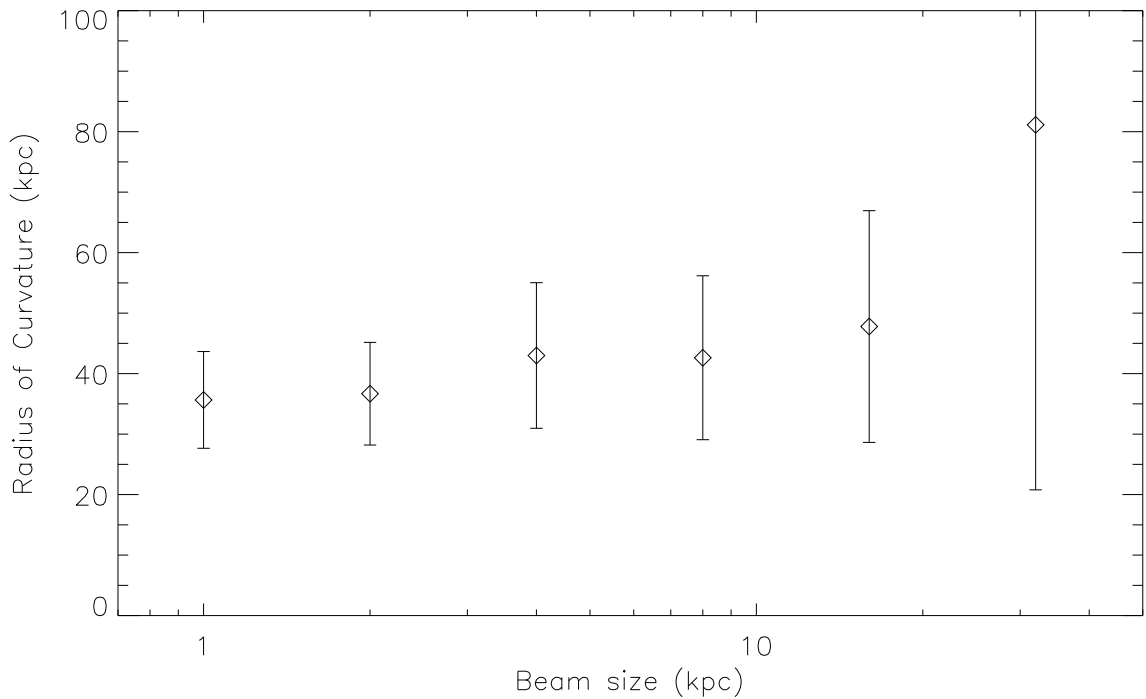}
\caption{
Measured radius of curvature vs. beam size for model .25E.  Error bars represent the overall error in measuring R.  As the beam size increases, the measured R initial remains fairly constant, but then begins to increase, along with the error, as the beam size approaches R.  Once the beam size is approximately equal to R, the radius cannot be reliably measured.
}
\label{fig:curve_vs_beam_size}
\end{figure*}

Inadequate resolution can also affect the measurement of the radius of curvature.  To characterize this, we apply a Gaussian smoothing filter to our radio images of simulation .25E to simulate radio beam FWHM sizes from 1~kpc to 32~kpc, and then used our fitting routine to find the radius of curvature.  The results are plotted in Fig.~\ref{fig:curve_vs_beam_size}.  For this simulation, the actual radius of curvature is $R=36$~kpc and 
the jet thickness is $h=2$~kpc.  For a marginally resolved jet (beam size $\leq 2$~kpc) the measured value of $R$ does not change.
For larger beam sizes, $R$ is over-estimated by about 10\% to 25\%.  
As the beam size becomes comparable to the radius of curvature, the fit for $R$ becomes very poor (error comparable to $R$).
Even in this case, however, the measured value of $R$ is systematically larger than the real value.
An over-estimate of $R$ will lead to an underestimate of the density derived from eqn.~\ref{eqn:r_over_h}, typically about 20\% for sources with unresolved jets.

Combining the effects of over-estimating the jet thickness and over-estimating the radius of curvature, we find that for a typical source in \citet{Free11} (resolved source, unresolved jet, $h/r \approx 1/4$) the density calculated is low by about 50\%, assuming a true value of $h/R$ of $1/17$.  This ranges from no correction for source S3 (marginally resolved jet) to about 85\% low for source S7 (beam size $\sim R$).  


For estimates of the jet luminosity, the error due to resolution can be significantly larger.  From eqn.~\ref{eqn:L_fit_pres}, we find that $L_{\rm jet} \propto h^{2} P_{\rm jet} \propto h^{6/7}$.  There is no dependance on $R$, but luminosity is very sensitive to $h$.
For example, assuming a real value of $h/R$ of $1/17$, the derived luminosity in table~\ref{table:sources_emily} would be reduced by a factor ranging from 1.2 (S3) to 13 (S7).


\subsection{Effect of Viewing Angles \label{sect:viewing_angle}}

So far, we have produced synthetic images of our simulations assuming that both the direction of jet propagation and the direction of motion of the AGN relative to the IGM are perpendicular to the observer's line of sight.  In reality, for observed bent-double radio sources there will be unknown angles between the jet direction and the observer and the direction of motion and the observer, both of which will affect the measurement of the radius of curvature.
The angle between the jet and the direction of motion will change where the point of maximum curvature is along the jet, but will not change the radius of curvature at that point \citep[e.g.][]{Begel79}.

\begin{figure*}
\includegraphics[scale=1.]{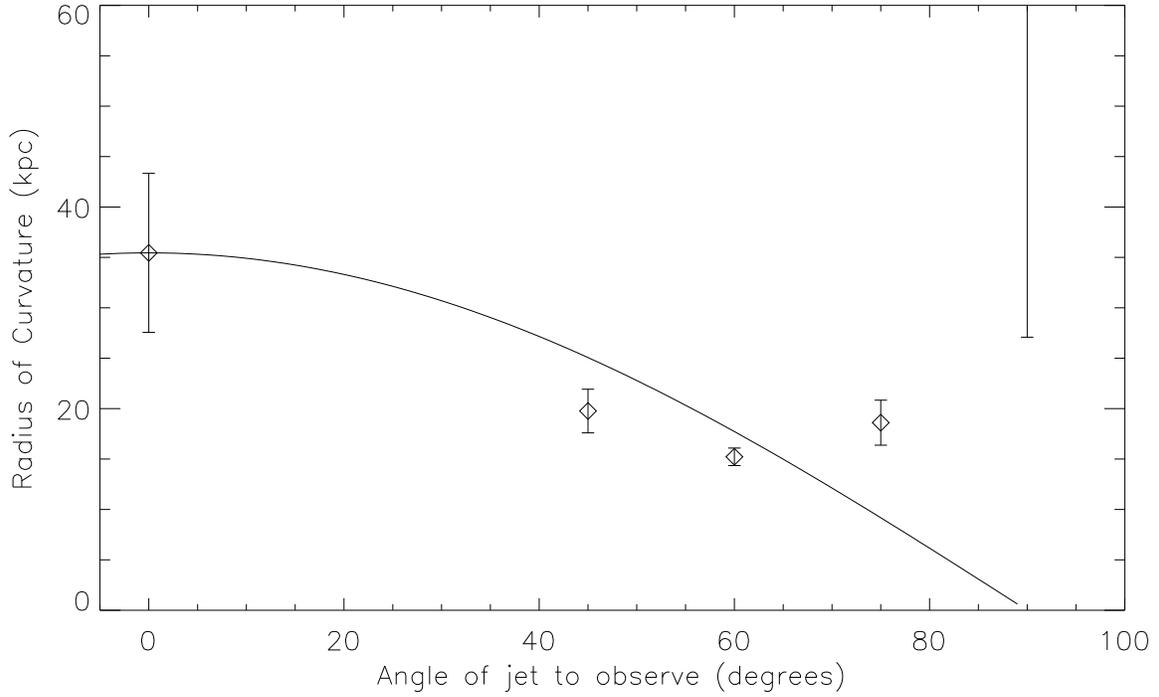}
\caption{
Measured radius of curvature vs. angle of jet direction relative to the observer ($0^{\circ}$ is perpendicular) for simulation .25E.  Solid line is $R = R_{(\theta=0)} \times \cos(\theta)$, the expected geometric correction.  The measured R follows this line fairly well out to about $60^{\circ}$, beyond which the radius of curvature is poorly fit.
}
\label{fig:curve_vs_angle_jet}
\end{figure*}

To quantify how much error viewing angle introduces, we take the data from simulation .25E, rotate it by various angles, then use it to create synthetic radio images.  We then measure the radius of curvature in each of these images by the procedure in sect.~\ref{sect:measure_r}.  
Fig.~\ref{fig:curve_vs_angle_jet} plots the measured radius of curvature for an angle between the jet direction and the observer between $0\degr$ and $90\degr$.  In all cases the direction of motion of the AGN is perpendicular to the observer's line of sight.  The solid line in Fig.~\ref{fig:curve_vs_angle_jet} is $R = R_{\rm (\theta=0)} \times \cos(\theta)$, which is the expected apparent radius of curvature of a simple rotated circle.  
The measured value of $R$ follows this line fairly well out to about $60\degr$, where the apparent radius is half the real radius.  
Beyond this, the measured radius does not get any smaller.  This is because the radio lobes at the ends of the jets begin to overlap from the observer's point of view, and their size dominates the fitting routine.
This is one possible explanation for the appearance of source S7 in \citet{Free11}, which appears to have a small radius of curvature but with very thick jets.
At $90\degr$, the fitting routine fails.  However, it is unlikely that a source with the jet aimed almost directly at the observer would be classified as a bent-double radio source.

\begin{figure*}
\includegraphics[scale=1.]{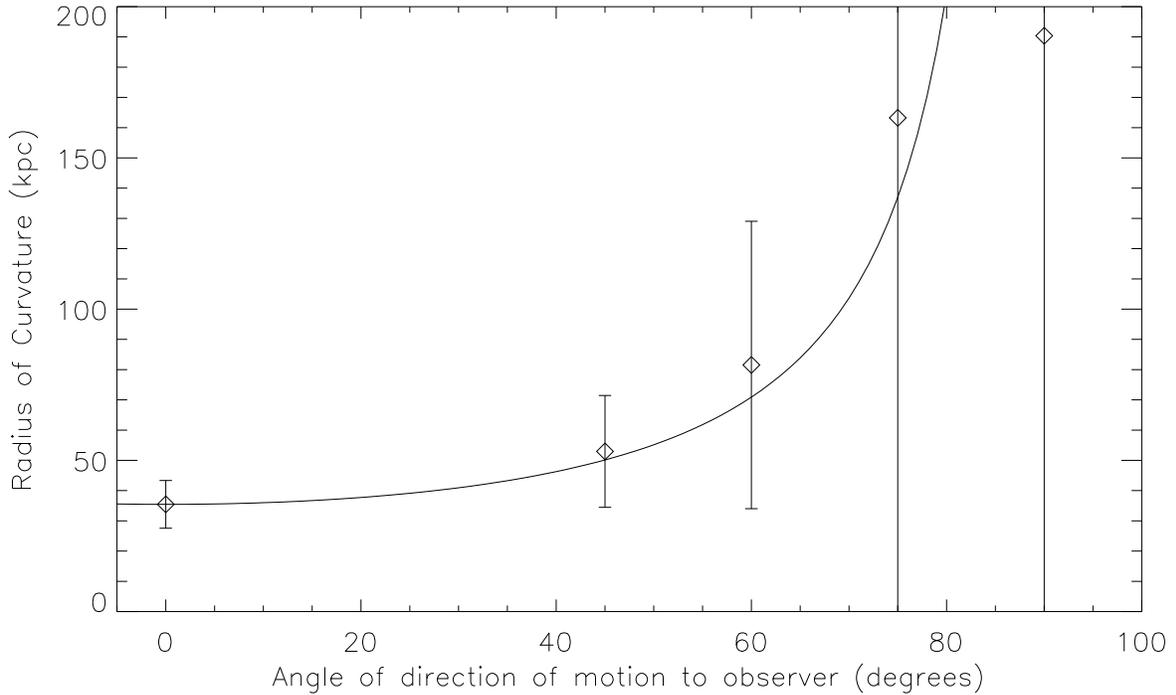}
\caption{
Measured radius of curvature vs. angle of direction of motion relative to the observer ($0^{\circ}$ is perpendicular) for simulation .25E.  Solid line is $R = R_{(\phi=0)} / \cos(\phi)$, the expected geometric correction.  The measured R follows this line fairly well out to about $60^{\circ}$, beyond which the radius of curvature is poorly fit.
}
\label{fig:curve_vs_angle_motion}
\end{figure*}

Fig.~\ref{fig:curve_vs_angle_motion} plots the measured radius of curvature for an angle between the direction motion of the AGN and the observer between $0\degr$ and $90\degr$.  In all cases the jet direction is perpendicular to the observer.  The solid line is Fig.~\ref{fig:curve_vs_angle_motion} is $R = R_{\rm (\phi=0)} / \cos(\phi)$, which is the expected apparent radius of curvature of a simple circle rotated in the same manner.  The measured value of $R$ follows this line out to about $60\degr$, where the apparent radius is double the real radius.
Beyond this, the source is so inclined that the jet appears the be approximately straight and the fitting routine fails.  


In practice, sources will be rotated around both axes, and orientation effects tend to cancel each other out somewhat.  Assuming sources rotated beyond $60\degr$ in either direction will not be classified as bent-double radio sources, the unknown viewing angle still introduces a large uncertainty in the measurement of $R$, and therefore in the values of $n_{\rm IGM}$ and $L_{\rm jet}$.  
On average, the error in the measurement of $R$ due to viewing angle should be between 50\% low to 30\% high ($1$-$\sigma$), but for an individual source could be up to a factor of 2 (100\% under-estimated to 50\% over-estimated).

\subsection{X-ray Detectability}

%
In general, the gas in galaxy groups is too cool and too low density to be seen in X-ray observations.  However, as an AGN moves through the IGM, a cocoon of shocked material develops around the jets.  The shock heats and compresses the IGM, boosting the X-ray emission.  The parameters that affect the X-ray brightness are the velocity of the AGN, $v_{\rm gal}$, which determines the temperature of the shocked gas, and the density of the IGM, $n_{\rm IGM}$, which will determine the emissivity at a given temperature.  

\begin{figure*}
\includegraphics[scale=.8]{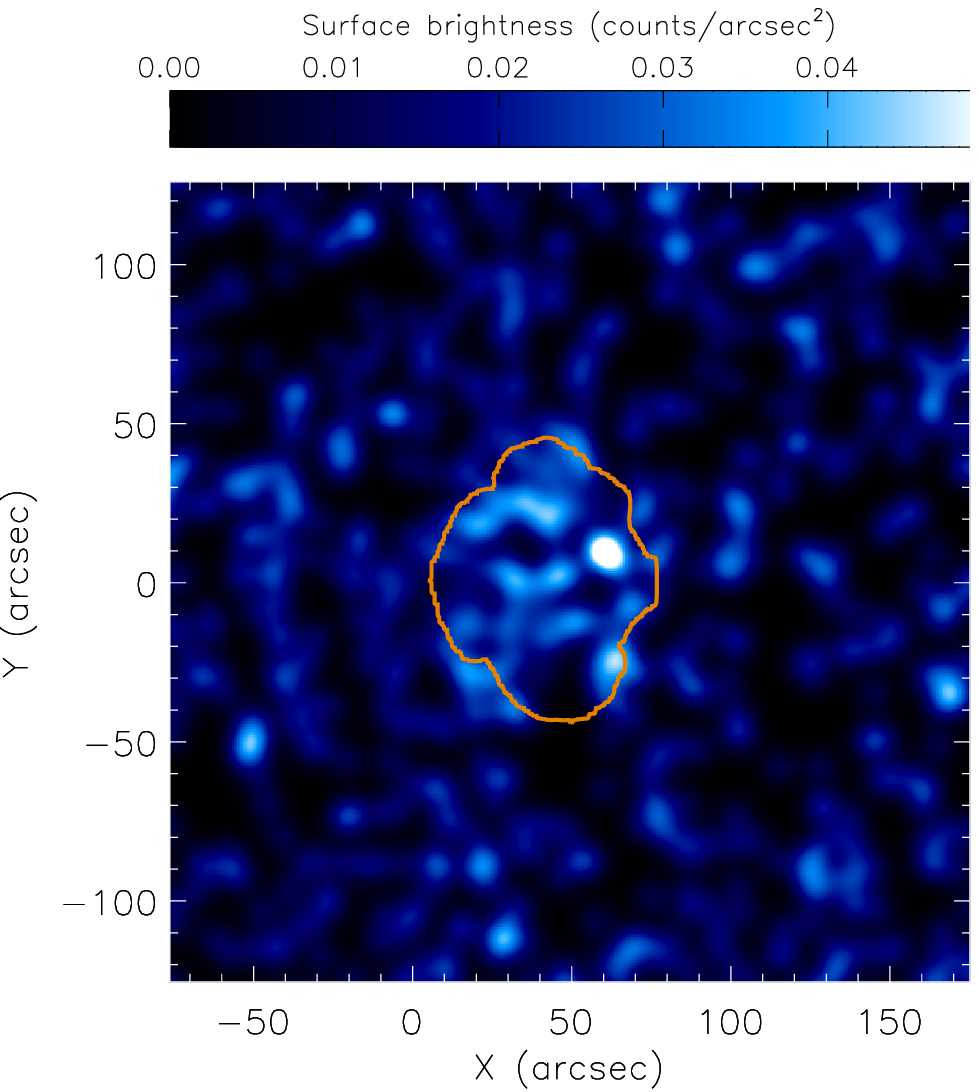}
\includegraphics[scale=.8]{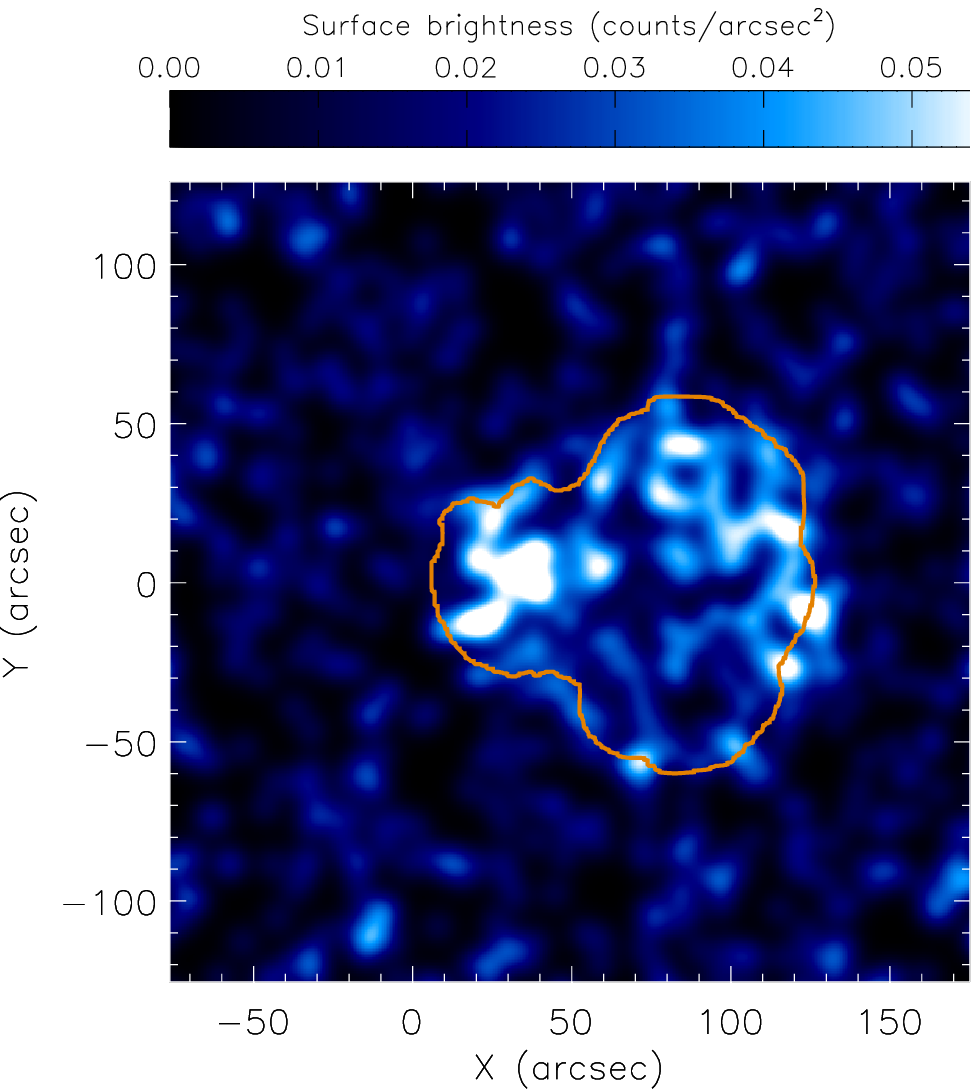}


\includegraphics[scale=.8]{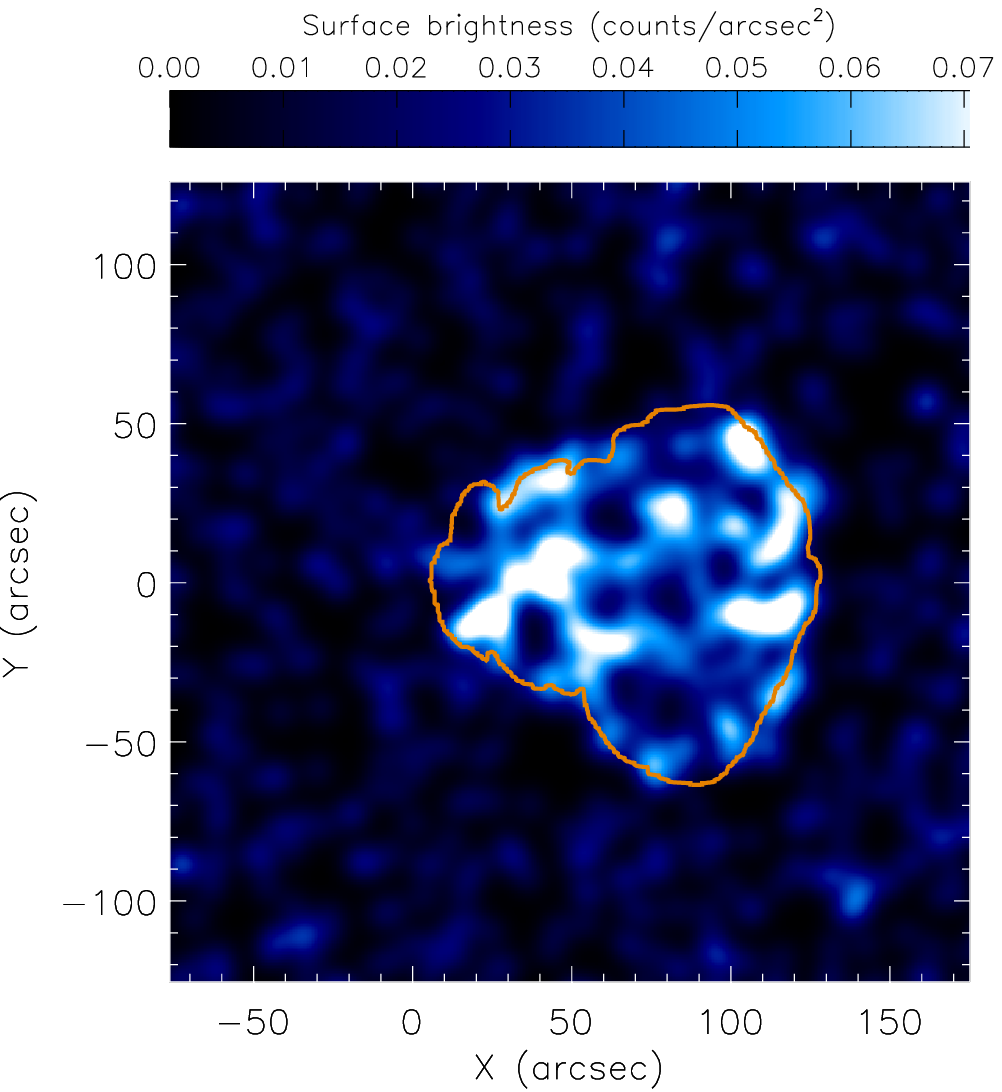}
\includegraphics[scale=.8]{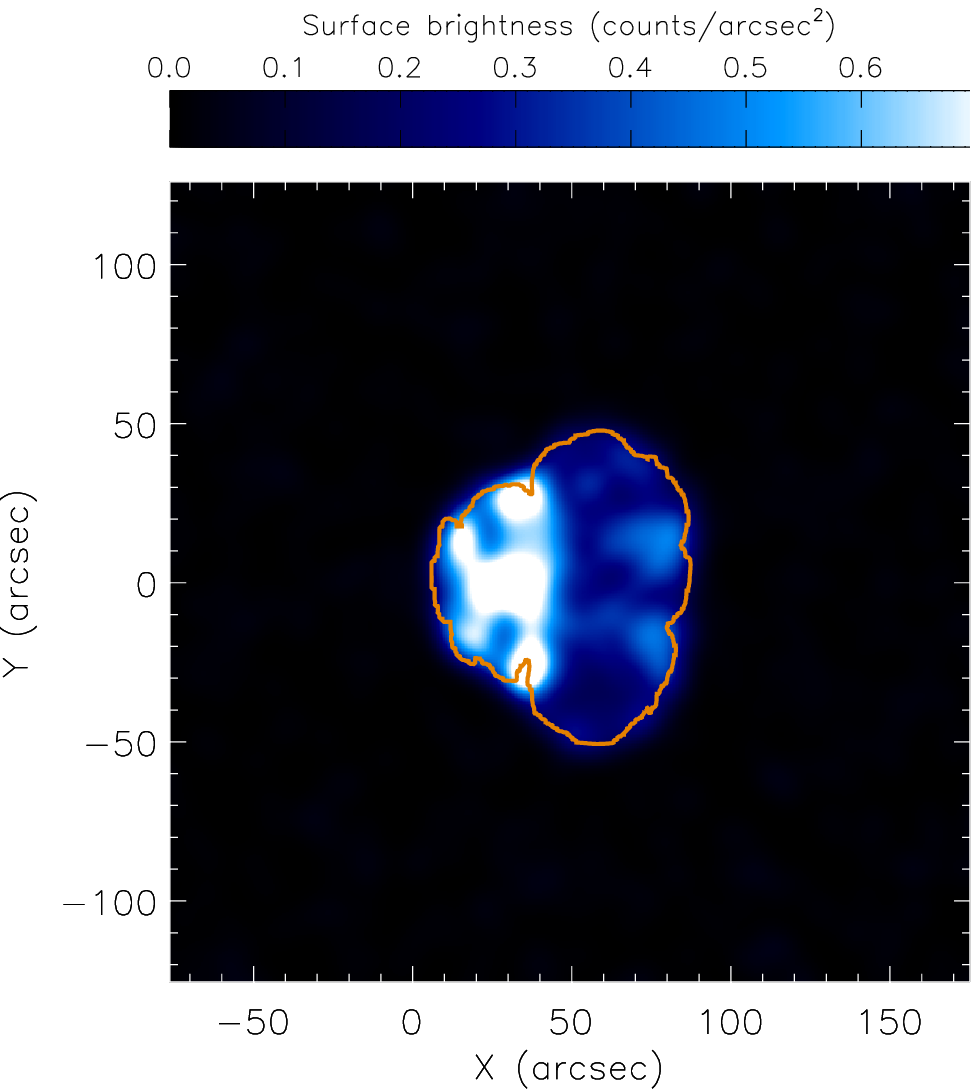}

\caption{
Simulated \textit{Chandra} X-ray observations for a 100ks of models .015E\_.25vel (upper left), .062E\_.5vel (upper right), .25E (bottom left) and 1E\_4n (lower right) at a redshift of $z = 0.1$.  All images are the total counts from $0.3$ to $3$~keV and are smoothed over 10 arcsec.  The radius of curvature of the jet in each model is about 36 kpc (20 arcsec).  Color bars show the scale of each image in counts/arcsec$^2$, scaled to $70\%$ of the maximum brightness in each image.  Orange contours mark the outer extent of radio emission.
}
\label{fig:xray_images_all}
\end{figure*}

To determine if bent-double radio sources in groups would be detectable in X-rays, we used the XIM tool \citep{Heinz09} to model 100~ks {\it Chandra} observations of four of our simulations, models .015E\_.25vel, .062E\_.5vel, .25E, and 1E\_4n, placed at a redshift of $z=0.1$.  Fig.~\ref{fig:xray_images_all} shows synthetic {\it Chandra} images.  In all figures we assume a background IGM temperature of $5\times10^5$~K and an IGM metallicity of $Z = 0.3$ solar.  All four simulations have about the same radius of curvature of $R\approx36$~kpc.  The first 3 simulations have the same IGM density with AGN velocities of $v_{\rm gal} = 250$~km/s (upper left), $500$~km/s (upper right) and $1000$~km/s (lower left).  The last two simulations have the same AGN velocity ($v_{\rm gal} = 1000$~km/s) but different densities of $n_{\rm IGM} = 10^{-3}$~cm$^{-3}$ (lower left) and $4\times10^{-3}$~cm$^{-3}$ (lower right).  For these images, we assume there is no point-source emission from the AGN.  Even if there is point-source emission, however, the high spatial resolution provided by {\it Chandra} would be able to remove this contribution.

Except in the lowest velocity case, there is significant X-ray emission from the region of the AGN jet and extended tail.  We would expect emission to be strongest at the leading edge of the jet, where we are seeing the strongest part of the shock edge-on.  However, because the edge-on shock is very thin and significant smoothing is needed to bring out the X-ray emission ($10$~arcsec in these images), the leading edge does not stand out. Instead, X-ray emission is spread across the region affected by the AGN and dominated by face-on shocks.  

\begin{figure*}
\includegraphics[scale=.8]{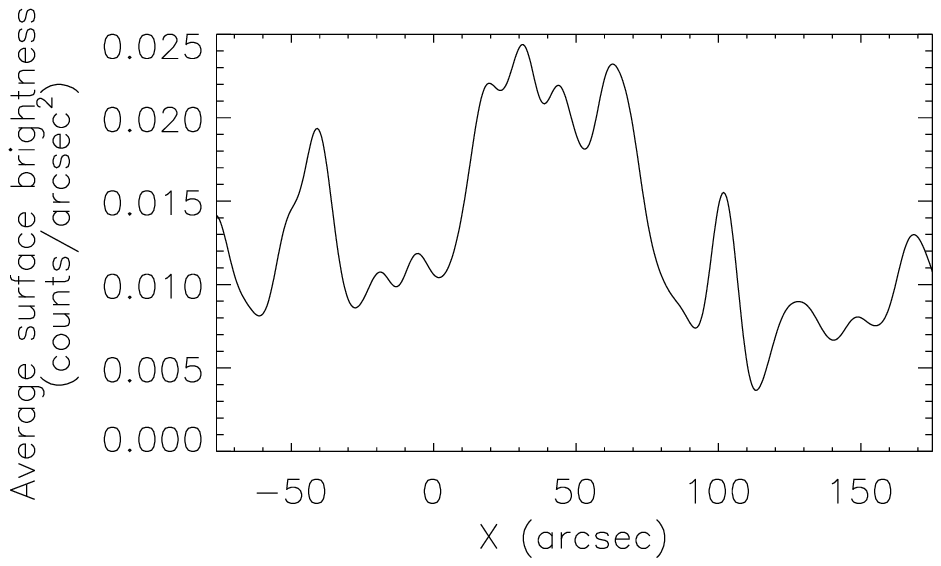}
\includegraphics[scale=.8]{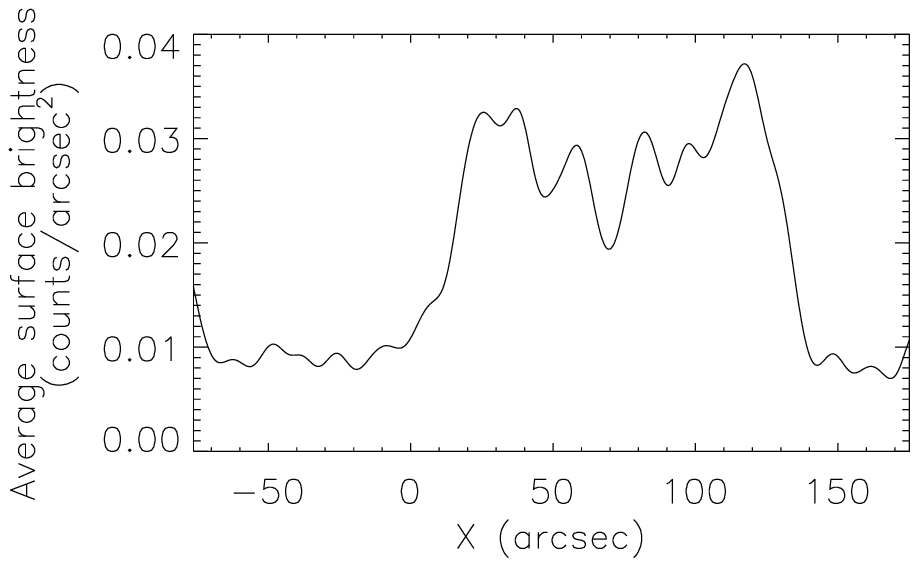}

\includegraphics[scale=.8]{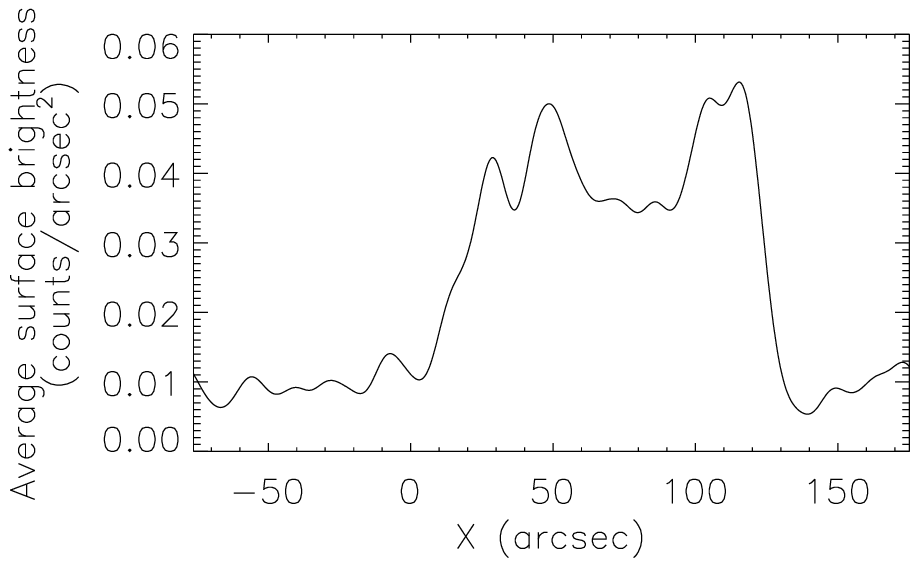}
\includegraphics[scale=.8]{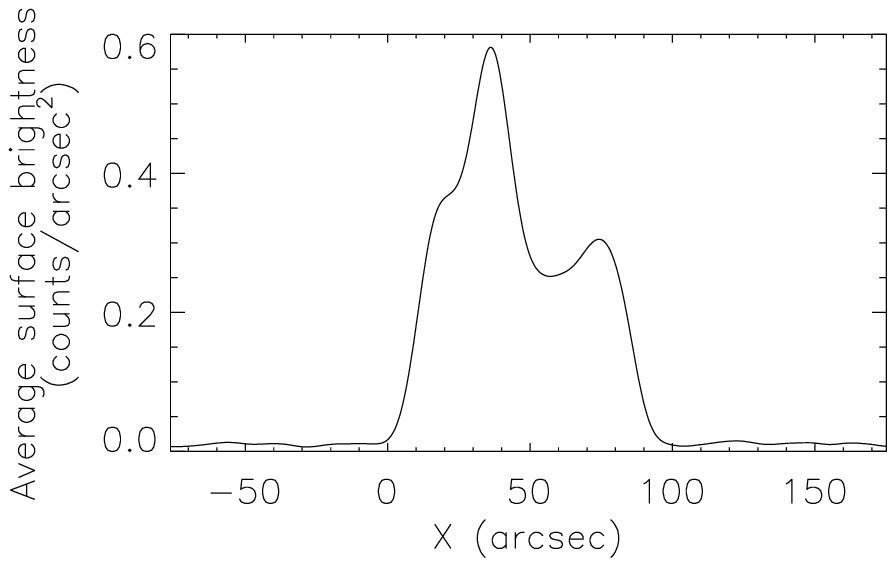}
\caption{
Plots of the mean surface brightness between $y = -50$ and $y = +50$ arcsec in each images in Fig.~\ref{fig:xray_images_all} vs. $x$ position.  Even in the worst case (model .015E\_.25vel, upper left), the mean surface brightness is about $.02$ counts/arcsec$^2$, about twice the background level.  This increases roughly as $v_{\rm gal}^{0.75}$ to about $.045$ counts/arcsec$^2$ for model .25E.  For model 1E\_4n, which has the same velocity as model .25E, the brightness increases to $0.3$ counts/arcsec$^2$, scaling as about $n_{\rm IGM}^{1.5}$.
}
\label{fig:xray_plots_all}
\end{figure*}

In Fig.~\ref{fig:xray_plots_all}, we plot the mean surface brightness between $-50$ and $+50$ arcsec of the AGN along the y-axis (parallel to the jet) in Fig.~\ref{fig:xray_images_all}.  In the worst-case presented here (model .015E\_.25vel, upper left panel), the surface brightness is about $0.02$~counts/arcsec$^2$ for a $100$~ks exposure, about twice the background level.  As the AGN velocity increases, the surface brightness increases as roughly $v_{\rm gal}^{\sim0.75}$, reaching about $0.045$ counts/arcsec$^2$ for model .25E.  For model 1E\_4n (lower right panel), which has the same velocity as model .25E, the brightness increases to $0.3$ counts/arcsec$^2$, a scaling of about $n_{\rm IGM}^{\sim1.5}$.  
In all cases, the shocks surrounding the AGN jet and tails are well resolved, so numerical mixing should have a minimal impact on the derived X-ray brightness.

If these models were placed at a higher redshift, the angular size of the X-ray source would decrease, but the surface brightness would remain the same.  For a large radius of curvature, the thickness of the shocked material, and therefore the surface brightness, will increase proportional to R.

X-ray observations of bent-double radio sources in groups would place important constraints on the IGM properties.  The X-ray surface brightness, along with an estimate of $v_{\rm gal}$ from velocity dispersion in the group, would allow $n_{\rm IGM}$ to be calculated independent of the radio observations.  Density calculated this way would also be less sensitive to the AGN velocity, scaling as about $n_{\rm IGM} \propto v_{\rm gal}^{\sim-0.5}$ rather than $n_{\rm IGM} \propto v_{\rm gal}^{-2}$ as in eqn.~\ref{eqn:r_over_h}.  The different scaling also means that in cases where $v_{\rm gal}$ is unconstrained, X-ray and radio data can be combined to find both the IGM density and AGN velocity.  This can also be used to refine the vales of $v_{\rm gal}$ and $n_{\rm IGM}$ and constrain the viewing angle in cases where there is an estimate for the AGN velocity.  
Because the X-ray emission follows the tail of extended radio emission, measuring the length of the X-ray emitting region would provide an age estimate for the source.

The general formula for the X-ray surface brightness is 
%
\begin{eqnarray}
S \simeq 0.1 \times (Z+0.1) \times v_{\rm gal,1000}^{0.75} \times n_{\rm -3}^{1.5} \times ( R/36 \textrm{~kpc} ) \nonumber \\
 \textrm{~counts/arcsec}^2
\label{eqn:xray_brightness}
\end{eqnarray}

\noindent for a 100~ks observation, where $Z$ is the metallicity of the IGM relative to solar. For sources S1 and S2 in table \ref{table:sources_emily} there are {\it Chandra} observations \citep{Free08} which found a total of $80\pm35$ and $94\pm26$ counts above the background in a 35.17~ks and 47.19~ks exposure, respectively.  Assuming the total area on the sky of these sources is about $4R^2$ (the actual size of the emitting region is unknown), the total counts predicted would be very roughly 30 counts each for $Z=0.3$.
This is lower than the observed counts, but current observations are not good enough to distinguish counts from the region of the bent-double from rest of the IGM, so they are still consistent.

For particularly bright sources, there may be enough photons to determine the temperature of the shock.  The temperature will scale roughly as $T \propto v_{\rm gal}^2$, so if the temperature can be found it would allow $v_{\rm gal}$ and $n_{\rm IGM}$ to be estimated from the X-ray data alone.  With additional constraints from radio and velocity dispersion observations, it would then be possible to find the compression ratio of the shock, and from that the temperature of the unshocked intra-group medium.


\section{Conclusions}




Our simulations are able to closely match the appearance of observed bent-double radio sources with narrow, curved jets ending in bright radio lobes.  We also predict that there should be an extended tail of radio emission that may be observable at low resolution and/or low frequencies.  The length of this tail would allow the age of the AGN to be determined.

From our simulations, we derive a formula for the radius of curvature (eqn.~\ref{eqn:radius_fit}) in terms of IGM density, AGN velocity, jet luminosity and jet velocity.  
%
%
From this we are able to calculate the kinetic jet luminosity (eqn.~\ref{eqn:L_fit_pres}) for observed bent-double radio sources, and find that $L_{\rm jet}$ is typically around $10^{45}$~erg/s, assuming that $v_{\rm jet} \simeq c$ and that the jets are really as thick as their observed values (see table~\ref{table:sources_emily}).  
The luminosities should be considered upper limits, as they will be lower if either the jets are slower or the jets are intrinsically thinner.
This formula is independent of $v_{\rm gal}$ and therefore can be used to find $L_{\rm jet}$ even when the velocity of the AGN is unknown.


A lack of resolution in radio observations leads to a systematic under-estimate of the IGM density for two reasons: the jet is unresolved and the radius of curvature is over-estimated.
In our simulations, use initial conditions that produce ratio for jet thickness to radius of curvature of $h/R \simeq 1/17$, equivalent to a $20\degr$ initial opening angle of the jet.  
For observed sources (limited by resolution), this ratio ranged from $1/14$ to $1/0.83$ in \citet{Free11}.  
Density scales as $(h/R)^{-1/7}$, so this leads to an under-estimate of the IGM density of up to $50\%$.
Inadequate resolution also leads to an over-estimate of the radius of curvature (and corresponding under-estimate of density), probably by about $20\%$.  
Overall, IGM density estimates for typical sources in \citet{Free11} are low by about $50\%$ due to resolution effects.  This ranges from no correction for source S3 (assuming a marginally resolved jet) to about 85\% low for source S7 (beam size $\sim R$).

The largest source of error, however, comes from the unknown angles between the observer, the jet direction and direction of motion of the AGN.  This can lead to either an over- or under-estimate of the true radius of curvature.  
This leads to an uncertainty of up to a factor of 2 in the estimate of the IGM density, with a typical ($1$-$\sigma$) error or about $50\%$.  
This is comparable to the error from all observational uncertainties (dominated by the uncertainty in the AGN velocity).
Even this uncertainty, however, does not change the conclusion that bending can be used to diagnose $\rho_{\rm IGM}$ in a statistical sense.

Finally, we have modeled the X-ray emission of bent-double radio sources and predict that they should be detectable in {\it Chandra} X-ray observations.  Although the IGM in groups is generally too cool and diffuse to be seen in X-ray observations, the shocks around the jet and tail in our simulations compress and heat the IGM, potentially making the entire region affected by the AGN jets detectable in X-rays.  The X-ray surface brightness scales as approximately $S \propto n_{\rm IGM}^{1.5} v_{\rm gal}^{0.75}$.
Sources in fairly dense environments and with fairly large angular size should be detectable in moderate ($\sim100$~ks) observations.  Count rates from existing short X-ray observations of sources S1 and S2 in \citet{Free08} are consistent with our predictions.  Further X-ray observations would provide complimentary constraints on the IGM density and AGN velocity to radio observations.

For particularly bright sources, it may be possible to obtain a measure of the temperature of the shock from X-ray observations.  This would provide an independent measure of the AGN velocity relative to the IGM. 

Future X-ray and radio observations will be able to place better constraints on the IGM density, AGN velocity and AGN age.  With very complete observations it should also be possible to constrain the temperature of the IGM and the orientation of the jets and direction of AGN motion.

\section*{Acknowledgments}

BJM is supported by an NSF Astronomy and Astrophysics Postdoctoral Fellowship under award AST1102796.
BJM and SH acknowledge NSF grant AST0707682.
SH acknowledges NSF grant AST1109347.
MB acknowledges support by the research group FOR 1254 funded by the Deutsche Forschungsgemeinschaft (DFG).
MR acknowledges NSF grant 1008454.
The software used in this work was in part developed by the DOE NNSA-ASC OASCR Flash Center at the University of Chicago.
Resources supporting this work were provided by the NASA High-End Computing (HEC) Program through the NASA Advanced Supercomputing (NAS) Division at Ames Research Center.

\bibliographystyle{mn2e}
\bibliography{references}


\end{document}